\documentclass{article}
\usepackage{spconf,amsmath,graphicx}
\usepackage{bm,amssymb,multirow,caption} 
\graphicspath{{pics/}}

\usepackage{enumitem}
\usepackage{xcolor}
\usepackage{fancyhdr}
\usepackage{atbegshi,picture}
\usepackage{lipsum}
\setlist{nosep, leftmargin=14pt}

\usepackage{mwe} 

\title{Dynamic Cardiac MRI Reconstruction using Combined Tensor Nuclear Norm and Casorati Matrix Nuclear Norm Regularizations}
\name{Yinghao Zhang, Yue Hu\thanks{This work is supported by the National Natural Science Foundation of China under Grant 61871159 and Natural Science Foundation of Heilongjiang YQ2021F005.}}
\address{School of Electronics and Information Engineering, Harbin Institute of Technology, Harbin, China}

\AtBeginShipoutNext{\AtBeginShipoutUpperLeft{%
  \put(\dimexpr\paperwidth-21cm\relax,-0.5cm){
    \footnotesize 
    This paper has been accepted in ISBI 2022 conference.}%
}}
\AtBeginShipoutNext{\AtBeginShipoutUpperLeft{%
  \put(\dimexpr\paperwidth-21cm\relax,-0.8cm){
    \footnotesize 
    Y. Zhang and Y. Hu, "Dynamic Cardiac MRI Reconstruction Using Combined Tensor Nuclear Norm and Casorati Matrix
    }%
}}
\AtBeginShipoutNext{\AtBeginShipoutUpperLeft{%
  \put(\dimexpr\paperwidth-21cm\relax,-1.1cm){
    \footnotesize 
    Nuclear Norm Regularizations," 2022 IEEE 19th International Symposium on Biomedical Imaging (ISBI), 2022, pp. 1-4, doi:
    }%
}}
\AtBeginShipoutNext{\AtBeginShipoutUpperLeft{%
  \put(\dimexpr\paperwidth-21cm\relax,-1.4cm){
    \footnotesize 
    10.1109/ISBI52829.2022.9761409.
    }%
}}

\begin{document}

\maketitle

\thispagestyle{fancy}
\cfoot{\footnotesize © 2022 IEEE. Personal use of this material is permitted. Permission from IEEE must be obtained for all other uses, in any current or future media, including reprinting/republishing
this material for advertising or promotional purposes, creating new collective works, for resale or redistribution to servers or lists, or reuse of any copyrighted component of this
work in other works.}


\renewcommand{\headrulewidth}{0mm}
\begin{abstract}
Low-rank tensor models have been applied in accelerating dynamic magnetic resonance imaging (dMRI). Recently, a new tensor nuclear norm based on t-SVD has been proposed and applied to tensor completion. Inspired by the different properties of the tensor nuclear norm (TNN) and the Casorati matrix nuclear norm (MNN), we introduce a combined TNN and Casorati MNN regularizations framework to reconstruct dMRI, which we term as TMNN. The proposed method simultaneously exploits the spatial structure and the temporal correlation of the dynamic MR data. The optimization problem can be efficiently solved by the alternating direction method of multipliers (ADMM). In order to further improve the computational efficiency, we develop a fast algorithm under the Cartesian sampling scenario. Numerical experiments based on cardiac cine MRI and perfusion MRI data demonstrate the performance improvement over the traditional Casorati nuclear norm regularization method.

\end{abstract}
\begin{keywords}
Dynamic MRI reconstruction, tensor nuclear norm, t-SVD
\end{keywords}
\section{Introduction}
\label{sec:intro}

Dynamic magnetic resonance imaging (dMRI) is one of the most important non-invasive imaging modalities, which has found a wide range of applications in clinical practice. However, due to the physical limitations, it is usually challenging to obtain dynamic MR images with high spatiotemporal resolution within clinically acceptable scan time. 
By exploiting the low-rank structure of the spatiotemporal Casorati matrix formulated by extracting and unfolding each time frame as a column, the low-rank matrix recovery methods \cite{ref_ktslr} have been successfully applied to reconstruct dynamic MR images from highly undersampled $k$-space data to accelerate dMRI. 

Low-rank tensor priors have been introduced as powerful alternatives due to their improvement in recovered image quality \cite{ref_tensor,ref_tucker1,ref_wmnn}. 
Compared with matrix, tensor is a more natural representation for multi-frame dynamic MR data. Recently, a new tensor decomposition called tensor singular value decomposition (t-SVD) \cite{ref_tsvd} has been proposed. Compared with the traditional tensor decompositions (e.g., CP, TUCKER), t-SVD can be easily computed by solving the matrix SVDs in the Fourier domain. Based on t-SVD, Lu \emph{et al.} proposed a new tensor nuclear norm (TNN) \cite{ref_tnn1}, which is the convex envelope of the tensor average rank. In addition, unlike certain traditional tensor constraints, utilizing the TNN constraint avoids explicit selection of tensor ranks. Some works have adopted this framework to reconstruct dMRI. Banco \emph{et al.} \cite{ref_tnn_dmri1} have applied t-SVD in dMRI reconstruction on a specific sampling mask by randomly and uniformly selecting $k$-space rows. TNN and total variation (TV) regularizations are combined in \cite{ref_tnn_dmri2} to improve the reconstruction of dMRI. 

Inspired by the distinct properties of the TNN and Casorati matrix nuclear norm (MNN), we introduce a novel dMRI reconstruction method using combined TNN and Casorati MNN regularizations, named as TMNN. Benefited from the combined framework, the proposed TMNN can simultaneously exploit the spatial structure and the temporal correlations of the dynamic MR image. The optimization problem can be efficiently solved by the alternating direction method of multipliers (ADMM). In order to further improve the computational efficiency, we introduce a novel fast algorithm to solve the TMNN minimization problem in the Cartesian sampling cases. We investigate the performance of the proposed method on the undersampled reconstruction of both cardiac cine MRI and myocardial perfusion MR data. 

\section{Notations and Preliminaries}
\label{sec:Notations}

In this paper, we denote tensors by Euler script letters, e.g., $\mathcal{X}$, matrices by bold capital letters, e.g., $\mathbf{X}$, vectors by bold lowercase letters, e.g., $\mathbf{x}$, and scalars by lowercase letters, e.g., $x$. For a 3-way tensor $\mathcal{X} \in \mathbb{C}^{n_1 \times n_2 \times n_3}$, we denote $\mathbf X_i$ as the $i$th frontal slice $\mathcal{X}(:,:,i),i=1,2,...,n_3$.

In \cite{ref_tsvd}, the t-SVD of $\mathcal{X} \in \mathbb{C}^{n_1 \times n_2 \times n_3}$ is formulated as 
\begin{equation}
\mathcal{X}=\mathcal{U} * \mathcal{S} * \mathcal{V}^{*}
\end{equation}
where $\mathcal{U}, \mathcal{V}, \mathcal{S} \in \mathbb{C}^{n_{1} \times n_{1} \times n_{3}}$, $\mathcal{S}$ is an $f$-diagonal tensor in which every frontal slice is diagonal, and the operator $*$ is the tensor-tensor product. Note that the formulation of t-SVD is similar to the matrix SVD. Based on the definition of t-SVD, TNN \cite{ref_tnn1} is defined by 
\begin{equation}
  \label{tnn}
\|\mathcal{X}\|_{*}=\frac{1}{n_{3}}\|\operatorname{bcirc}(\mathcal{X})\|_{*}=\frac{1}{n_{3}}\|\bar{\mathbf X}\|_{*}
\end{equation}
where $\operatorname{bcirc}(\mathcal{X})$ represents the block circulant matrix of $\mathcal{X}$:
\begin{equation}
  \label{bcirc}
  \operatorname{bcirc}(\mathcal{X})=
  \left[
    \begin{array}{cccc}
    \mathbf X_1& \mathbf X_{n_3} & \cdots & \mathbf X_2 \\
    \mathbf X_2& \mathbf X_1 & \cdots & \mathbf X_3 \\
    \vdots & \vdots & \ddots & \vdots \\
    \mathbf X_{n_3} & \mathbf X_{n_3-1} & \cdots & \mathbf X_1
    \end{array}
  \right]
\end{equation}
In (\ref{tnn}), $\bar{\mathbf X}$ denotes the block diagonal matrix:
\begin{equation}
  \label{bdiag}
\bar{\mathbf X}=\operatorname{bdiag}(\bar{\mathcal{X}})=\left[\begin{array}{llll}
\bar{\mathbf X}_1 & & & \\
& \bar{\mathbf X}_2 & & \\
& & \ddots & \\
& & & \bar{\mathbf X}_{n_3}
\end{array}\right]
\end{equation}
where $\bar{\mathcal{X}}=\operatorname{fft}(\mathcal{X},[],3)$ denotes the result of DFT on $\mathcal{X}$ along the third dimension.

\section{Methods}
\label{sec:Methods}

\subsection{The proposed TMNN model}
\label{subsec:model}

We denote the  distortion-free dynamic MR image as $\mathcal{X} \in \mathbb{C}^{n_{1} \times n_{2} \times n_{3}}$, where $n_1$, $n_2$ denote the spatial coordinates, and $n_3$ is the temporal coordinate. The $i$th frontal slice $\mathbf X_i$ is the $i$th frame of dMRI. The data acquisition of dMRI can be modeled as
\begin{equation}
  \mathbf b = A(\mathcal{X}) +\mathbf{n}
\end{equation}
where $\mathbf b \in \mathbb{C}^{m}$ is the observed undersampled $k$-space data, $A: \mathbb{C}^{n_{1} \times n_{2} \times n_{3}} \rightarrow \mathbb{C}^m$ is the Fourier sampling operator, and $\mathbf{n} \in \mathbb{C}^{m}$ is the Gaussian distributed white noise.

According to \cite{ref_tensor}, $\mathcal{X}$ can be unfolded into three different matrices, namely, mode-$n$ unfolding, $n=1,2,3$. The mode-1 and mode-2 unfoldings preserve the spatial image structure of each time frame but are unable to capture the structure of temporal profiles. TNN can be calculated by the nuclear norm of the block circulant matrix \eqref{bcirc}, of which each block row and column are the mode-1 and mode-2 unfolding of the tensor, respectively. Therefore, similar to the combination of nuclear norms of mode-1 and mode-2 unfoldings, TNN is capable of utilizing the spatial information at each time frame but unable to take full advantage of the correlations between different time evolutions. Different from the mode-1 and mode-2 unfoldings, the mode-3 unfolding, also known as Casorati matrix, arranges the temporal profiles of each spatial voxel into rows, thereby preserving the time evolution. However, since each time frame is vectorized as a column in the Casorati matrix, the spatial structure of the data can not be fully captured. 

Inspired by the different properties of the TNN and the Casorati MNN, we propose a novel algorithm using the combined TNN and Casorati MNN regularizations, which we term as TMNN. The optimization problem can be formulated as follows
\begin{equation}
  \small
\label{model}
\min_{\mathcal{X}} \frac12 {\Vert A(\mathcal{X})-\mathbf b \Vert}_F^2+\lambda_1{\Vert \mathcal{X} \Vert}_*+\lambda_2{\Vert \mathbf C (\mathcal{X}) \Vert}_*
\end{equation}
where ${\Vert \mathcal{X} \Vert}_*$ is the tensor nuclear norm of ${\cal X}$, $\mathbf C: \mathbb{C}^{n_{1} \times n_{2} \times n_{3}} \rightarrow \mathbb{C}^{n_{1} n_{2} \times n_{3}}$ unfolds the tensor into a Casorati matrix, ${\Vert \mathbf C (\mathcal{X}) \Vert}_*$ denotes the nuclear norm of the Casorati matrix $\mathbf C(\mathcal{X})$, and $\lambda_1$, $\lambda_2$ are the regularization parameters.

\subsection{Optimization Algorithm Based on ADMM}
\label{classical alg}

The optimization problem \eqref{model} can be solved based on the following variable splitting algorithm:
\begin{equation}
  \small
  \label{VS}
  \begin{aligned}
    &\min_{\mathcal{X}} \frac12 {\Vert A(\mathcal{X})-\mathbf b \Vert}_F^2+\lambda_1{\Vert \mathcal{Z} \Vert}_*+\lambda_2{\Vert \mathbf M \Vert}_* \\
    &s.t. \quad \mathcal{Z} = \mathcal{X},\  \mathbf M = \mathbf C (\mathcal{X})
  \end{aligned}
  \end{equation}
The equality constrained problem \eqref{VS} can be efficiently solved with the alternating direction method of multipliers (ADMM) algorithm, resulting in the following iterative scheme:
\begin{footnotesize}
\begin{align}
  &\mathcal{Z}^{n+1}=\arg \min_{\mathcal{Z}} \lambda_1{\Vert \mathcal{Z} \Vert}_*+ \left \langle \mathcal{W}_1^{n}, \mathcal{Z}-\mathcal{X}^{n} \right \rangle + \frac{\mu_1}2 \Vert \mathcal{Z}-\mathcal{X}^{n} \Vert_F^2 \label{Z} \\
  &\begin{aligned}
    \mathbf M^{n+1}=\arg\min_{\mathbf M} \lambda_2{\Vert \mathbf M \Vert}_* &+ \left \langle \mathbf{W}_2^{n}, \mathbf M-\mathbf C (\mathcal{X}^{n}) \right \rangle \\
     &+ \frac{\mu_2}2 \Vert \mathbf M-\mathbf C (\mathcal{X}^{n}) \Vert_F^2 \label{M} 
  \end{aligned} \\
  &\begin{aligned}
    \mathcal{X}^{n+1}=\arg\min_{\mathcal{X}} \frac12 {\Vert A(\mathcal{X})-\mathbf b \Vert}_F^2 &+ \left \langle \mathcal{W}_1^{n}, \mathcal{Z}^{n+1}-\mathcal{X} \right \rangle \\
    + \frac{\mu_1}2 \Vert \mathcal{Z}^{n+1}-\mathcal{X} \Vert_F^2& + \left \langle \mathbf{W}_2^{n}, \mathbf M^{n+1}-\mathbf C (\mathcal{X}) \right \rangle \\
    &+ \frac{\mu_2}2 \Vert \mathbf M^{n+1}-\mathbf C (\mathcal{X}) \Vert_F^2 \label{x1}
  \end{aligned} \\
  &\mathcal{W}_1^{n+1} = \mathcal{W}_1^{n}+\mu_1(\mathcal{Z}^{n+1}-\mathcal{X}^{n+1})\\
  &\mathbf W_2^{n+1} = \mathbf W_2^{n}+\mu_2(\mathbf M^{n+1}-\mathbf C (\mathcal{X}^{n+1}))
\end{align}
\end{footnotesize}

\noindent where $\mathcal{W}_1, \mathbf{W}_2$ are Lagrange multipliers. For simplification, we omit the superscript in the following description.

The subproblems \eqref{Z} and \eqref{M} can be solved by the tensor singular value thresholding (TSVT) \cite{ref_tnn1} and the SVT, respectively. The subproblem \eqref{x1} is a quadratic problem and hence can be solved analytically as, 
\begin{equation}
  \footnotesize
    \mathcal{X} =\frac{A^* \mathbf b+\mu_1\mathcal{Z}+\mu_2 \mathbf C^* \mathbf M+\mathcal{W}_1+\mathbf C^* \mathbf{W}_2}
    {A^*A+\mu_1 \mathbf 1+\mu_2\mathbf 1}
\end{equation}
In the above equation, the division is performed element-wise, and $\mathbf 1$ denotes an all-one tensor.
If the measurements are $k$-space samples on a Cartesian grid, $\mathcal{X}$ can be efficiently computed. In that case, $A=\mathcal{S}\mathcal{F}$, where $\mathcal{S}$ is the under-sampling mask and $\mathcal{F}$ transforms the dMRI image into $k$-space. Thus, we can obtain
\begin{equation}
  \footnotesize
  \label{x2}
  \mathcal{X} = \mathcal{F}^* \left[\frac{\mathcal{S}^*\mathbf b+\mathcal{F} \left(\mu_1\mathcal{Z}+\mu_2 \mathbf C^* \mathbf M+\mathcal{W}_1+\mathbf C^* \mathbf{W}_2\right)}{\mathcal{S}^* \mathcal{S}+\mu_1\mathbf 1+\mu_2\mathbf 1}\right]
\end{equation}

\subsection{Fast optimization algorithm}
\label{fast alg}

Based on the properties of TNN and the Casorati MNN, we develop a computationally efficient algorithm for the proposed TMNN method in the case of Cartesian sampling. 

For a Casorati dMRI matrix $\mathbf{X} \in \mathbb{C}^{n_1  n_2 \times n_3}$, we can obtain the $k$-space data $\hat{\mathbf X}$ by,
\begin{equation}
  \hat{\mathbf{X}} = (\mathbf F_1 \otimes \mathbf F_2) \mathbf X
\end{equation}
where $\mathbf F_1$ and $\mathbf F_2$ are the Hermite symmetric FFT over the first dimension (column) and the second dimension (row), respectively, and the operator $\otimes$ is the Kronecker product. It is easy to prove that $(\mathbf F_1 \otimes \mathbf F_2)^*(\mathbf F_1 \otimes \mathbf F_2)$ is equal to identity matrix. 
As well known, the matrix nuclear norm of $\mathbf X$ equals to the sum of its singular values, which are obtained by the eigenvalues of $\mathbf X^* \mathbf X$. Thus, we can obtain the following relation, 
\begin{equation}
  \label{proof_mnn}
\hat{\mathbf{X}}^*\hat{\mathbf{X}} = \mathbf X^*(\mathbf F_1 \otimes \mathbf F_2)^*(\mathbf F_1 \otimes \mathbf F_2) \mathbf X = \mathbf X^* \mathbf X
\end{equation}
According to the property above, the Casorati MNN satisfies
\begin{equation}
  \label{mnn_p}
\Vert  \mathbf{X} \Vert_* = \Vert  \hat{\mathbf{X}} \Vert_*
\end{equation}

For a 3D dMRI tensor $\mathcal{X} \in \mathbb{C}^{n_1 \times n_2 \times n_3}$, we can obtain each frontal slice $\hat{\mathbf{X}}_i$ of $\hat{\mathcal{X}}$ in the $k$-space by,
\begin{equation}
  \hat{\mathbf{X}}_i = \mathbf F_1 \mathbf{X}_i \mathbf F_2,\;i=1,2,...,n_3
\end{equation}
According to \eqref{tnn} and \eqref{bdiag}, we can obtain
\begin{equation}
  \footnotesize
\begin{aligned}
&\left\Vert 
  \begin{array}{llll}
  \bar{\mathbf X}_1 & & & \\
  & \bar{\mathbf X}_2 & & \\
  & & \ddots & \\
  & & & \bar{\mathbf X}_{n_3}
  \end{array}\right\Vert_*   \\
= &\left\Vert 
  \begin{array}{llll}
    \mathbf F_1 \bar{\mathbf X}_1 \mathbf F_2 & & & \\
    &\mathbf F_1 \bar{\mathbf X}_2 \mathbf F_2& & \\
    & & \ddots & \\
    & & & \mathbf F_1 \bar{\mathbf X}_{n_3} \mathbf F_2
  \end{array}\right\Vert_* \\
=&\Vert \hat{\mathcal{X}} \Vert_*
\end{aligned}
\end{equation}
Thus, the TNN based on t-SVD decomposition satisfies
\begin{equation}
  \label{tnn_p}
  \Vert \mathcal{X} \Vert_* = \Vert \hat{\mathcal{X}} \Vert_*
\end{equation}
Based on \eqref{mnn_p} and \eqref{tnn_p}, when the undersampled measurements are collected on the Cartesian grid, we can rewrite the reconstruction model \eqref{model} as a simple tensor completion optimization problem,
\begin{equation}
  \label{model2}
  \min_{\hat{\mathcal{X}}} \frac12 {\Vert S \hat{\mathcal{X}}-\mathbf b \Vert}_F^2+\lambda_1{\Vert \hat{\mathcal{X}} \Vert}_*+\lambda_2{\Vert \mathbf C(\hat{\mathcal{X}}) \Vert}_*
\end{equation}
The optimization problem above can be solved using similar ADMM steps as shown in \ref{classical alg}, except for the $\hat{\mathcal{X}}$ subproblem, which can be solved analytically as,
{\small
\begin{equation}
  \label{x2_}
  \hat{\mathcal{X}} = \frac{\mathcal{S}^*\mathbf b+\mu_1\mathcal{Z}+\mu_2 \mathbf C^* \mathbf M+\mathcal{W}_1+\mathbf C^* \mathbf{W}_2}{\mathcal{S}^* \mathcal{S}+\mu_1\mathbf 1+\mu_2\mathbf 1}
\end{equation}}

Compared with \eqref{x2}, computations of the FFTs are not required in (22) in the fast algorithm, which substantially reduces the computational complexity. Instead of reconstructing the spatial dynamic MR image from the undersampled $k$-space data, we directly reconstruct the $k$-space MR data in the Fourier domain, which is able to decrease the computational burden in solving the nuclear norm constrained minimization problem. 
Note that the way we convert the dynamic MRI reconstruction into a simple tensor completion problem can be easily generalized to other optimization problems. 


\section{Experimental results}
\label{experiments}

We evaluate the performance of the proposed TMNN method based on two data, i.e., a cardiac cine MR image with the size of $256\times 256\times 10$ and a myocardial perfusion MR image with the size of $190\times 90\times 70$. We assume that the measurements are acquired using the pseudo radial Cartesian sampling \cite{ref_ktslr} and variable density random sampling patterns under different undersampling ratios. We also add complex Gaussian white noise with the signal-to-noise ratio (SNR) of 20dB to the undersampled $k$-space data. The balancing parameters are experimentally set to be $\lambda_1=2.5e^{-3}$ and $\lambda_2=7.5e^{-3}$ for the noiseless case, and $\lambda_1=\lambda_2=0.1$ for the noisy case. 

\begin{figure}[t]
\centering
{\footnotesize
\leftline{\qquad  Fully sampled \qquad \qquad MNN recovery  \qquad \qquad   TMNN recovery }
}
\centerline{\includegraphics[width=8.5cm]{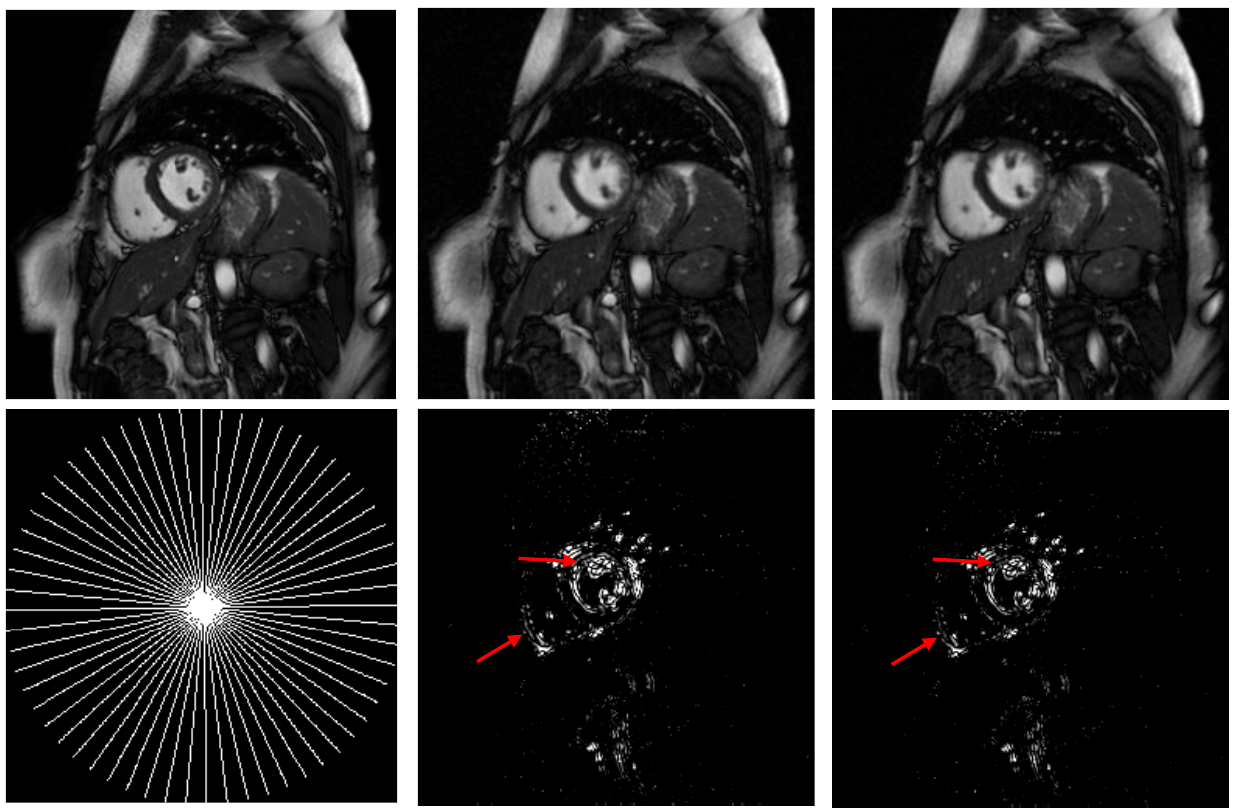}}
\caption{Recovery of the cine MRI data from the noiseless radial undersampled measurements using TMNN. The error images are shown in the second row.}
\label{fig1}
\end{figure}

\begin{figure}[t]
\centering
{\footnotesize
\leftline{\qquad  Fully sampled \qquad \qquad MNN recovery  \qquad \qquad   TMNN recovery }
}
\centerline{\includegraphics[width=8.5cm]{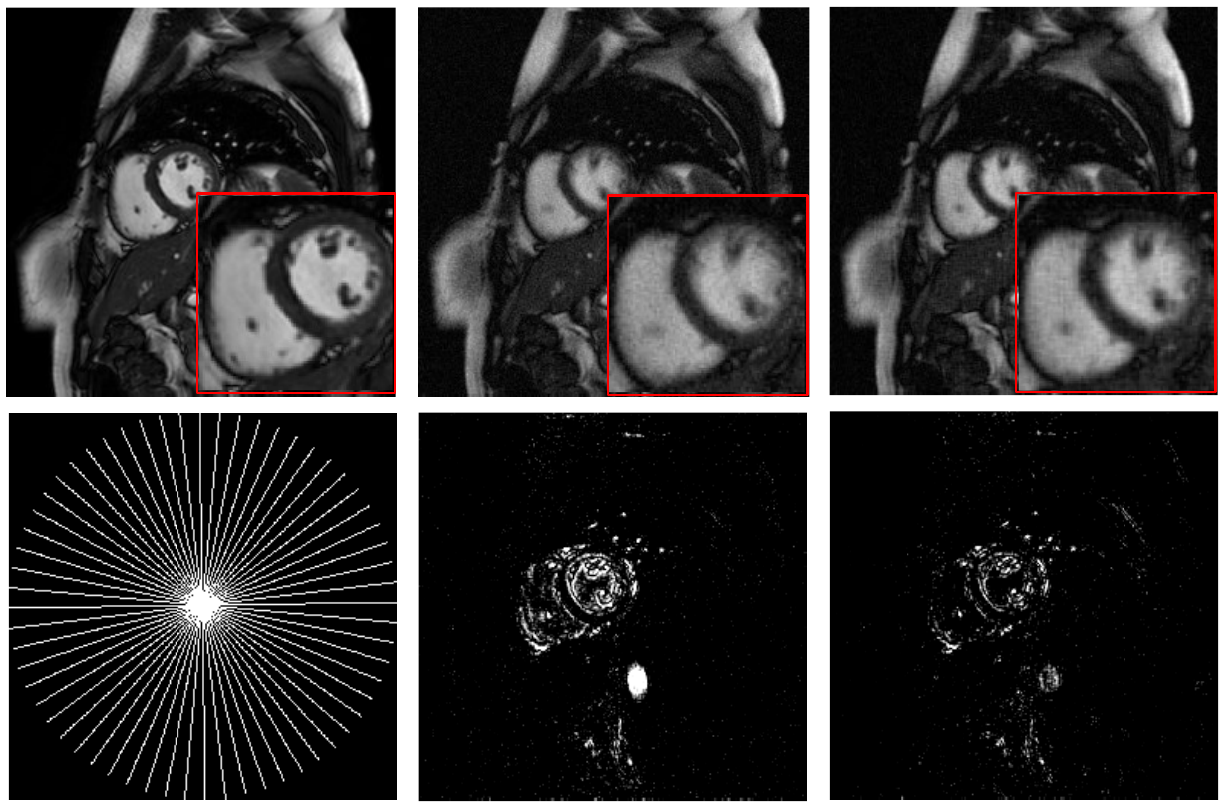}}
\caption{Recovery of the cine MRI data from the noisy radial undersampled measurements using MNN and TMNN. The error images are shown in the second row.}
\label{fig2}
\end{figure}

\begin{figure}[htb]
  \centering
{\footnotesize
\leftline{\qquad  Fully sampled \qquad \qquad MNN recovery  \qquad \qquad   TMNN recovery }
}
\centerline{\includegraphics[width=8.5cm]{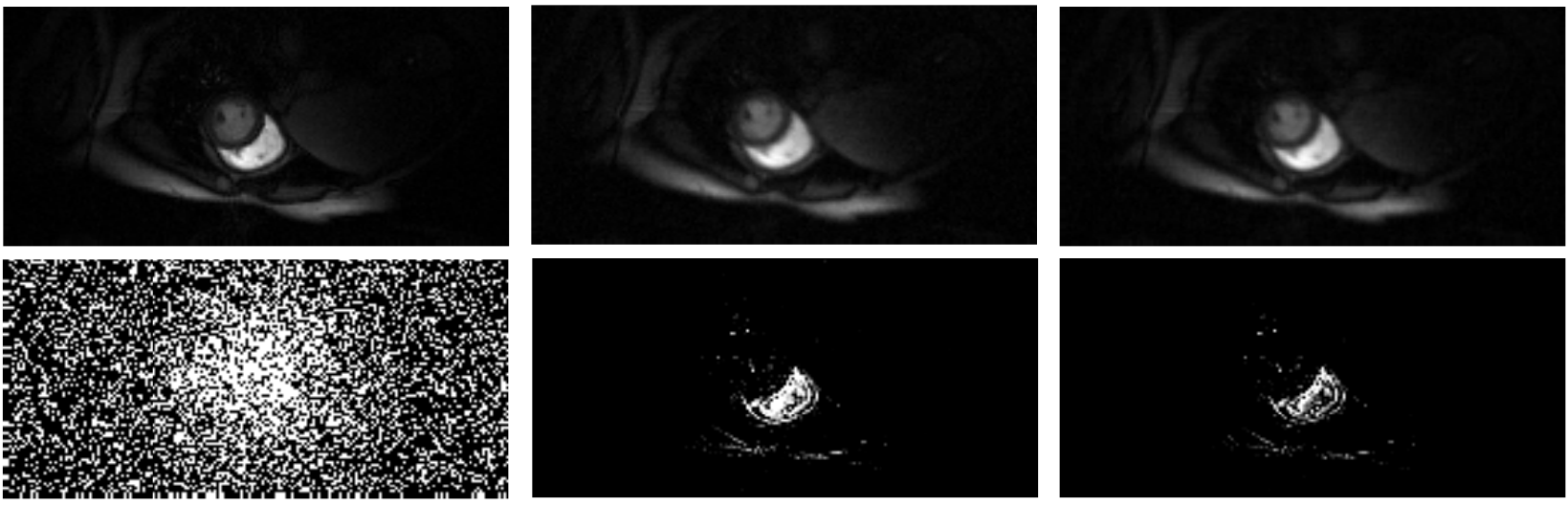}}
\caption{Recovery of the myocardial perfusion MRI data from the noisy variable density random undersampled measurements using MNN and TMNN. The error images are shown in the second row.}
\label{fig3}
\end{figure}

In Fig.\ref{fig1}, we compare the recovery results of the TMNN with MNN on a cine cardiac MR image from 30 radial lines (undersampling ratio $\sim$0.1) in the noiseless case. We observe that the proposed TMNN model outperforms the MNN method in providing more accurate reconstruction. Fig.\ref{fig2} shows the reconstruction of the cine cardiac MR image from the noisy undersampled measurements using 30 radial lines. In Fig.\ref{fig3}, we plot the noisy reconstruction results of the perfusion MR image from the variable density random sampling trajectory with the undersampling ratio of 0.3. It is observed that the TMNN method generates less error compared with the MNN approach. 
The SNRs of the reconstructed dynamic image using TNN, MNN, and the proposed TMNN at different undersampling conditions are shown in Table.\ref{tab1}. We observe that except for one case, the proposed TMNN consistently provides the best reconstruction results and improves the SNR by up to 2dB over the MNN method.
In addition, it is shown that the improvement of the proposed TMNN over MNN is more significant in the noisy setting.

In order to investigate the utility of the proposed fast algorithm in \ref{fast alg}, we compare the running time of the fast algorithm and the ADMM algorithm in \ref{classical alg}. For both algorithms, we terminate the iteration process till the relative change of the cost function is less than a predefined tolerance value. Experiments show that when the stopping condition is satisfied, the two methods obtain the same SNR, and the fast algorithm is eight percent faster on average. Particularly, the computational time for the algorithm in 3.3 and 3.2 on the cine MR image is 63.9s and 69.6s, respectively.

\begin{table}[h]
  \footnotesize
  \center
\caption{ SNR comparisons of TNN, MNN, and the proposed TMNN on two datasets using four sampling schemes.}
  \label{tab1}
  \begin{tabular}{cclllll}
    \hline
    \multicolumn{1}{l}{}                                                     & \multicolumn{1}{l}{}       &      & \multicolumn{2}{c}{Radial/ratio(lines)}                & \multicolumn{2}{c}{Random/ratio}                    \\ \cline{4-7} 
    \multicolumn{1}{l}{}                                                     & \multicolumn{1}{l}{}       &      & \multicolumn{1}{c}{0.05(16)} & \multicolumn{1}{c}{0.1(30)} & \multicolumn{1}{c}{0.1} & \multicolumn{1}{c}{0.3} \\ \hline
    \multirow{6}{*}{\begin{tabular}[c]{@{}c@{}}Cine\\ MRI\end{tabular}}      & \multirow{3}{*}{Noisy}     & TNN  & 13.12                  & 14.99                  & \textbf{16.74}          & 18.28                   \\
                                                                             &                            & MNN  & 11.14                  & 13.27                  & 16.56                   & 17.26                   \\
                                                                             &                            & TMNN & \textbf{13.35}         & \textbf{15.09}         & 16.63                   & \textbf{18.50}          \\ \cline{2-7} 
                                                                             & \multirow{3}{*}{Noiseless} & TNN  & 17.06                  & 19.78                  & 20.62                   & 24.16                   \\
                                                                             &                            & MNN  & 18.15                  & 20.22                  & 20.56                   & 23.55                   \\
                                                                             &                            & TMNN & \textbf{18.45}         & \textbf{21.02}         & \textbf{20.87}          & \textbf{24.71}          \\ \hline
    \multirow{6}{*}{\begin{tabular}[c]{@{}c@{}}Perfusion\\ MRI\end{tabular}} & \multirow{3}{*}{Noisy}     & TNN  & 12.60                  & 14.56                  & 14.09                   & 14.45                   \\
                                                                             &                            & MNN  & 12.80                  & 14.61                  & 14.41                   & 14.55                   \\
                                                                             &                            & TMNN & \textbf{13.65}         & \textbf{15.51}         & \textbf{14.95}          & \textbf{15.65}          \\ \cline{2-7} 
                                                                             & \multirow{3}{*}{Noiseless} & TNN  & 14.34                  & 16.94                  & 18.55                   & 17.77                   \\
                                                                             &                            & MNN  & 16.19                  & 18.10                  & 18.02                   & 19.00                   \\
                                                                             &                            & TMNN & \textbf{16.40}         & \textbf{18.43}         & \textbf{18.80}          & \textbf{19.46}          \\ \hline
    \end{tabular}
  \end{table}

\section{Conclusion}
\label{conclude}

We proposed a novel combined regularization algorithm for dMRI reconstruction. By combining the tensor nuclear norm and the Casorati matrix nuclear norm, both the low-rank properties of the Casorati matrix and the tensor can be captured to exploit the spatiotemporal structures of the dataset. In order to efficiently solve the proposed optimization problem, we adopt the ADMM algorithm. Moreover, we developed a faster algorithm to solve the minimization problem when the samples are collected using Cartesian trajectories, which further improve the computational efficiency. Experimental results demonstrate the improved performance of the proposed TMNN model over the low-rank matrix recovery method.





\bibliographystyle{IEEEbib}
{\footnotesize
\bibliography{refs}}

\end{document}